# Gate-reconfigurable Josephson microwave source limited by thermal noise

Jayeon Kim[1], Seong Jang[1,2,*], Seunghan Lee[1], Dongsung T. Park[1], Kenji Watanabe[3], Takashi Taniguchi[4], and Gil-Ho Lee[1,*]

[1] Department of Physics, Pohang University of Science and Technology, Pohang 37673, Republic of Korea

[2] Department of Electrical and Computer Engineering, Northeastern University, Massachusetts 02115, USA

[3] Research Center for Electronic and Optical Materials, National Institute for Materials Science, 1-1 Namiki, Tsukuba 305-0044, Japan

[4] Research Center for Materials Nanoarchitectonics, National Institute for Materials Science, 1-1 Namiki, Tsukuba 305-0044, Japan

[*]Correspondence to Seong Jang (seongjang1@postech.ac.kr) and Gil-Ho Lee (lghman@postech.ac.kr)

**ABSTRACT**

Josephson junctions are compact on-chip sources of coherent microwave and terahertz radiation, but conventional radiators couple the emission frequency, output power and dissipation through a single bias voltage. Here we demonstrate a gate-reconfigurable Josephson microwave radiator based on a graphene weak link integrated with a separately gated graphene shunt resistor. The bias voltage sets the Josephson frequency, while the local gates tune the Josephson current and the total device resistance, enabling electrostatic control of emitted power at a chosen frequency. The device resolves Josephson harmonics up to fourth order, and its radiation power scales with source current and impedance as expected from a lumped-element circuit model. We further show that the emission linewidth is quantitatively described by Johnson-Nyquist noise at the electron temperature. The linewidth therefore provides an on-chip thermometer under bias, with the resistance-dependent heating captured by a minimal heat-balance model based on Wiedemann–Franz diffusion and electron–phonon coupling.

Josephson junctions (JJs) are central building blocks of superconducting quantum technologies, serving as the nonlinear circuit elements that underpin qubits, parametric amplifiers, and quantum-accurate voltage standards [1-3]. A key functionality of JJs is the generation of coherent electromagnetic radiation via the alternating-current (AC) Josephson effect, which directly links a dc bias voltage to an emission frequency [4-6]. This makes JJs natural on-chip sources of microwave and terahertz radiation, with applications spanning quantum control, readout, and spectroscopy [7,8]. In conventional Josephson radiators [9], however, the emission frequency, output power, and dissipation are intrinsically coupled through the bias voltage. Tuning the emission frequency by adjusting the bias inevitably modifies the power delivered to the electromagnetic environment and the associated noise, precluding independent control over these quantities. This constraint limits the functionality of Josephson sources as reconfigurable circuit elements, and also obscures the physical mechanisms that govern their spectral properties — in particular, the role of thermal fluctuations and electron temperature in determining the emission linewidth. While linewidth narrowing has been demonstrated in specific regimes, such as cavity-coupled and flux-flow oscillators [8,10-12] and recent on-chip Josephson sources have integrated tunable emission with detection [13], independent electrostatic control of the critical current and circuit resistance — the two parameters that set the output power — within a single device has yet to be demonstrated. Gate-tunable Josephson junctions based on van der Waals materials and semiconductor weak links have emerged as a versatile platform for exploring the AC Josephson effect beyond conventional tunnel junctions [14-18]. In particular, graphene encapsulated in hexagonal boron nitride (hBN) provides a high-mobility, gate-tunable SNS weak link with high junction transparency, enabling a strongly non-sinusoidal current–phase relation and gate-controlled supercurrent [19-22]. These properties make graphene JJs attractive candidates for reconfigurable Josephson sources, yet their use as Josephson radiators — and in particular the spectral characterization of their emission — has remained largely unexplored.

Here we introduce a gate-reconfigurable Josephson radiator based on an hBN-encapsulated graphene weak link, in which the emission frequency and output power can be controlled largely independently using a dual-gate architecture that integrates a gate-tunable graphene JJ in parallel with a separately gated graphene shunt resistor. This design enables orthogonal control of frequency and power on a single chip, defining a reconfigurable operating space not accessible in conventional Josephson radiators. Within this reconfigurable operating space, we perform emission spectroscopy and analyze the linewidth of Josephson radiation. We find that the linewidth is quantitatively accounted for by Johnson-Nyquist noise at the electron temperature. Leveraging this relation, we use the linewidth as a direct probe of the electron temperature under bias and show that its dependence on the device resistance is accurately described by a minimal heat-balance model combining Joule heating with Wiedemann–Franz diffusion and electron–phonon coupling in the normal-metal electrodes [23,24]. Our results establish a graphene-based Josephson radiator with independently tunable frequency and power, and demonstrate that its spectral properties can be understood in terms of thermal noise. More broadly, this work introduces a reconfigurable platform for on-chip Josephson radiation and provides a route to quantitatively probe electron heating and cooling in mesoscopic conductors, with implications for the design of low-noise quantum microwave sources and nanoscale thermal devices.

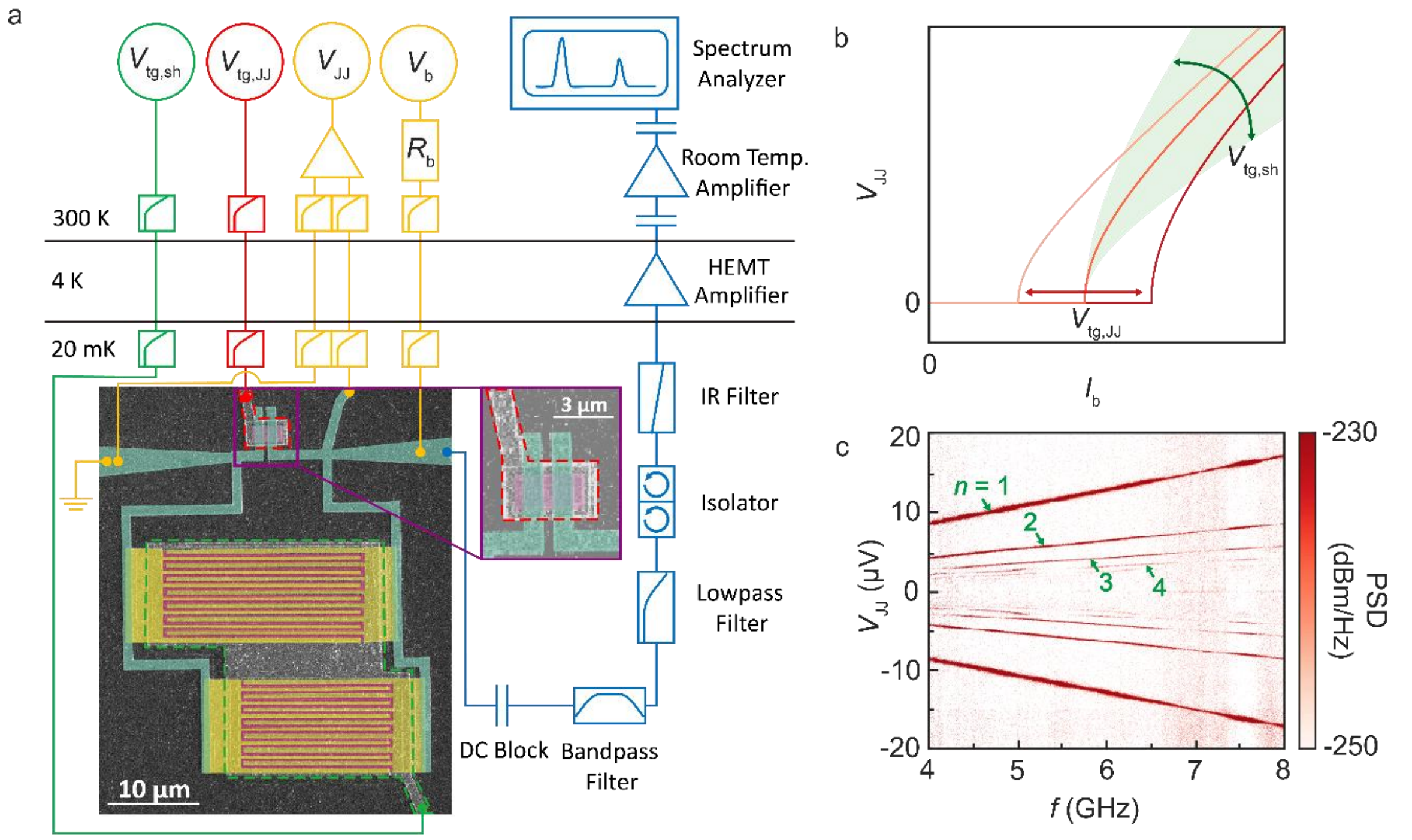

**Fig. 1 | Device structure and measurement of Josephson radiation.** (a) False-colored scanning electron microscope (SEM) image of the device overlaid with a schematic of the power spectral density (PSD) measurement setup. The graphene channel (pink) is encapsulated by hexagonal boron nitride (hBN) and contacted by molybdenum–rhenium (MoRe) superconducting electrodes (blue) and gold normal-metal electrodes (yellow). Bias current $I_b$ is applied by sourcing a voltage $V_b$ through a series resistance $R_b$, and the junction voltage $V_{JJ}$ is recorded simultaneously. Top gates for the Josephson junction ($V_{tg,JJ}$) and the shunt resistor ($V_{tg,sh}$) are indicated by the red and green dashed lines, respectively. The emitted radiation is collected through a cryogenic microwave chain, amplified by a high-electron-mobility transistor (HEMT) amplifier, and detected at room temperature with a spectrum analyzer. (b) $I_b$–$V_{JJ}$ characteristics illustrating independent gate control. Since the shunt resistance $R_{sh}$ is designed to satisfy $R_{sh} \ll R_N$, where $R_N$ is the normal-state resistance of the Josephson junction, the total resistance is dominated by $R_{sh}$. Varying $V_{tg,JJ}$ (red arrow) tunes the critical current $I_c$ while leaving the total resistance unchanged, whereas varying $V_{tg,sh}$ (green arrow) adjusts the total resistance (green shadowed region) without modifying $I_c$. (c) Josephson radiation spectra as a function of $V_{JJ}$ and detection frequency $f$, measured at $V_{tg,JJ}$ = 2 V and $V_{tg,sh}$ = 5 V. Green arrows indicate higher-order harmonics following the Josephson frequencies $f_n = n(2eV_{JJ})/h$, where $e$ is the electron charge, $h$ is Planck's constant, and $n$ = 1, 2, 3, 4 is the harmonic order.

## Gate-reconfigurable Josephson radiator

Our device consists of a graphene Josephson junction (GJJ) connected in parallel with a graphene normal-metal resistor acting as a shunt (Fig. 1a). The Josephson radiation frequencies are given by $f_n = n(2eV_{JJ})/h$, where $e$ is the electron charge, $h$ is Planck's constant, and $n$ is the harmonic order [4], and can therefore be controlled via the junction voltage $V_{JJ}$ across the device. The critical current $I_c$ and the total resistance $R_{total}$ are controlled independently by dedicated top gates applied to the junction and the shunt, respectively. This dual-gate architecture decouples the control of emission frequency from that of output power — a capability not accessible in conventional Josephson radiators, where both quantities are tied to the same bias voltage. This decoupling has a further practical consequence rooted in the condition $R_{sh} \ll R_N$, where $R_{sh}$ is the shunt resistance and $R_N$ is the normal-state resistance of the Josephson junction: because $R_{total}$ is dominated by $R_{sh}$, $V_{JJ}$ remains low at a given bias current $I_b$ [25,26]. This keeps the fundamental Josephson frequency $f_1$ within the microwave range accessible by standard instrumentation. Moreover, changes in $V_{tg,JJ}$ produce only a small variation in $V_{JJ}$, allowing $f_1$ to be held fixed by fine-tuning $V_{tg,sh}$ or $I_b$ while $I_c$ is varied independently (Fig. 1b). The emitted radiation spectrum confirms that the device operates as a Josephson radiator (Fig. 1c). Clear emission lines following the AC Josephson relation are observed across the full measurement bandwidth, with harmonics up to order $n$ = 4 prominently resolved. The strong higher-harmonic content reflects the highly non-sinusoidal current–phase relation of the junction [22,27-29], consistent with the high junction transparency expected for a superconductor–normal–superconductor (SNS) weak link based on hexagonal boron nitride (hBN)-encapsulated graphene [20,21,30].

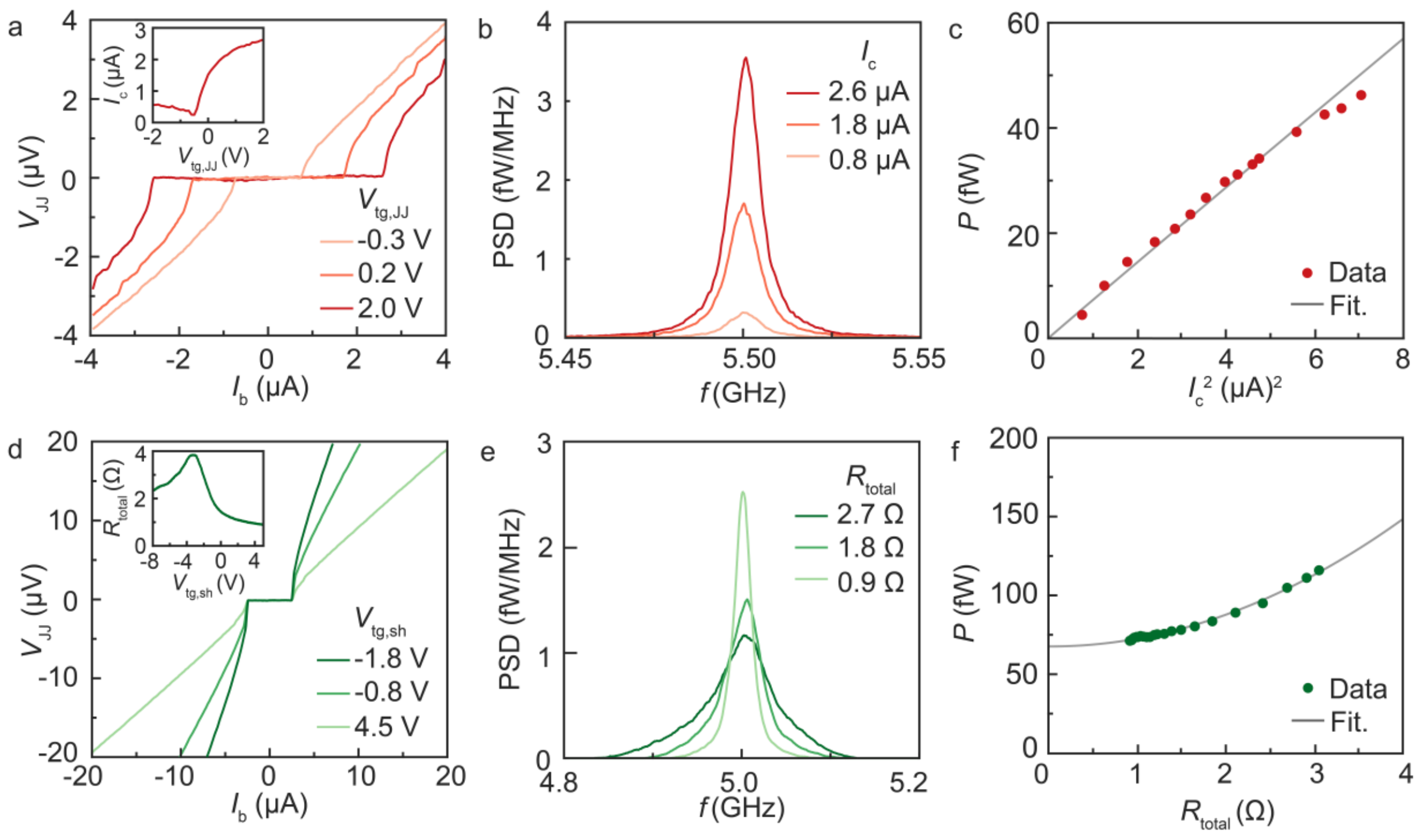

**Fig. 2 | Source-current and impedance-tunable Josephson radiation.** (a) $I_b$–$V_{JJ}$ characteristics for different values of $V_{tg,JJ}$. Inset: Extracted $I_c$ as a function of $V_{tg,JJ}$, where $I_c$ is defined as the bias current

$I_b$ at which $V_{JJ}$ exceeds 0.2 μV. (b) PSD as a function of $f$ for different values of $I_c$, after frequency-dependent background subtraction. (c) Radiation power $P$ delivered to the load (HEMT amplifier) as a function of $I_c^2$, obtained by integrating the PSD over frequency. Symbols represent data and the solid line is a linear fit, $P \propto I_c^2$. (d) $I_b$–$V_{JJ}$ characteristics for different values of $V_{tg,sh}$. Inset: total device resistance $R_{total}$ as a function of $V_{tg,sh}$. (e) PSD as a function of $f$ for different values of $R_{total}$, after frequency-dependent background subtraction. (f) Radiation power $P$ as a function of $R_{total}$, obtained by integrating the PSD. Symbols represent data and the solid line is a fit based on a power-division model.

## Electrostatic control of output power

We first demonstrate power control through the junction gate. Varying $V_{tg,JJ}$ modifies $I_c$ without altering $R_{total}$ (Fig. 2a), since $R_{sh} \ll R_N$ ensures that $R_{total}$ is set entirely by the shunt. As $I_c$ sets the amplitude of the Josephson current oscillations, varying $V_{tg,JJ}$ directly controls the amplitude of the emitted radiation [31]. The radiation spectra recorded at fixed $V_{JJ}$ (Fig. 2b) show that the emitted power increases with $I_c$ while $f_1$ remains unchanged. Integrating the power spectral density (PSD) over frequency, after correcting for the frequency-dependent gain of the measurement chain (see Supplementary Information Fig. S1), yields the total power $P$ delivered to the load (Fig. 2c), which follows $P \propto I_c^2$, consistent with the expected scaling of Josephson radiation [6,31]. We next demonstrate power control through the shunt gate. Varying $V_{tg,sh}$ tunes $R_{total}$ from approximately 1 to 3 Ω (Fig. 2d) without modifying $I_c$, and hence without changing the amplitude of the Josephson oscillations. Here, the output power is instead controlled by the impedance seen by the load: as $R_{total}$ is varied, the fraction of generated radiation delivered to the external load is redistributed accordingly [31]. The radiation spectra (Fig. 2e) show substantial variation in the emitted power at fixed $f_1$, and the integrated power (Fig. 2f) is well described by a power-division model (See Supplementary Information Fig. S2). A closer look at the radiation spectra in Figs. 2b and 2e reveals an intriguing pattern: while the linewidth remains largely unchanged as $I_c$ is varied, it broadens noticeably with increasing $R_{total}$. This asymmetry suggests that the linewidth is sensitive to the device resistance rather than the emitted power itself, pointing to a thermal noise origin [31].

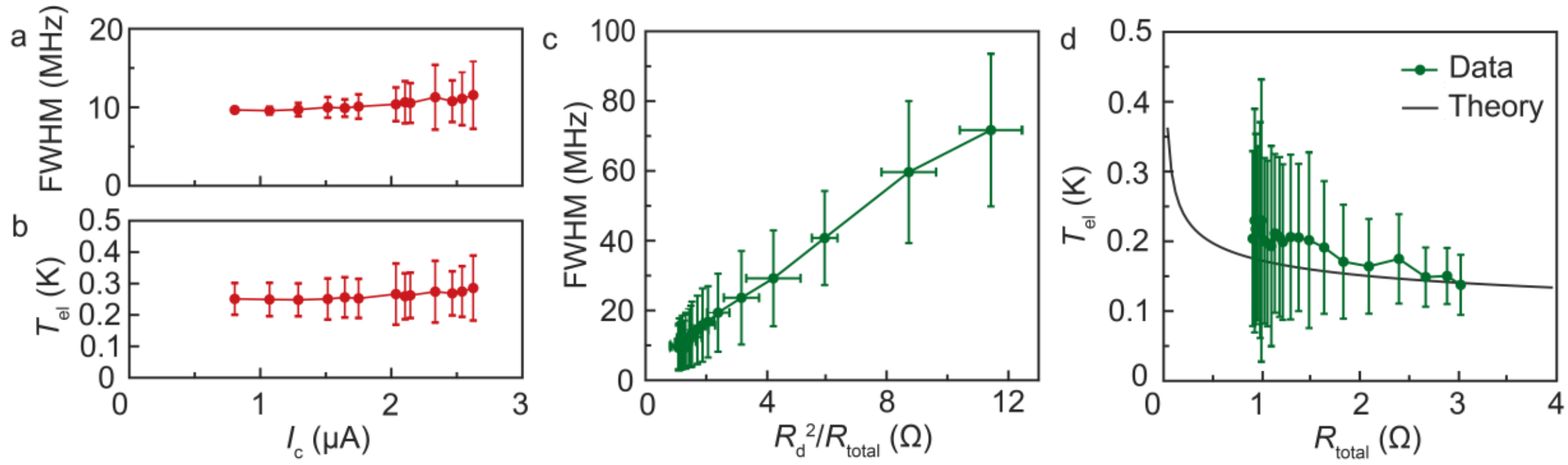

**Fig. 3 | Josephson radiation linewidth and electron temperature.** (a) Full width at half maximum

(FWHM) of the Josephson radiation as a function of $I_c$, tuned by $V_{tg,JJ}$, measured at $V_{tg,sh}$ = 3 V and $T_{bath}$ = 20 mK. Each data point represents the average FWHM of radiation peaks with emission frequencies ranging from 4.8 to 5.2 GHz, with error bars indicating the standard deviation. (b) Electron temperature $T_{el}$ extracted from the linewidth in (a), plotted versus $I_c$. (c) FWHM as a function of $R_d^2/R_{total}$, where $R_d$ is the differential resistance of the junction, measured at $V_{tg,JJ}$= 2 V and $T_{bath}$ = 20 mK. (d) $T_{el}$ as a function of $R_{total}$. Symbols represent experimental data, while the black solid curve is a heat-balance calculation, with $T_{el}$=110 mK at the bath temperature.

### Thermal-noise-limited linewidth

The spectral linewidth of Josephson radiation is determined by phase diffusion driven by current fluctuations in the circuit [32,33]. In the thermal limit, the dominant contribution is Johnson-Nyquist noise [34,35], which leads to a full width at half maximum (FWHM) of the emission peak given by FWHM = $4\pi(2e/h)^2R_d^2/R_{total}k_BT_{el}$, where kB is the Boltzmann constant and $T_{el}$ is the electron temperature of the resistive elements in the circuit, especially the shunt (see Supplementary Information Fig. S4 and Ref. [33,36]). This expression shows that the linewidth is governed by two quantities: the ratio $R_d^2/R_{total}$, controlled by $V_{tg,sh}$, and $T_{el}$, which is set by the heat balance of the electronic system under bias. We first vary $I_c$ via $V_{tg,JJ}$ while keeping $R_{total}$ fixed. The FWHM (Fig. 3a) and the corresponding $T_{el}$ extracted from the expression above (Fig. 3b) both remain unchanged across the full range of $I_c$. This is consistent with the condition $I_b \gg I_c$, under which $R_d \approx R_{total}$ and the linewidth becomes insensitive to $I_c$. We next vary $R_{total}$ via $V_{tg,sh}$ while keeping $I_c$ fixed. Figure 3c shows that the FWHM increases linearly with $R_d^2/R_{total}$, in quantitative agreement with the expression above. This confirms that the linewidth is governed by Johnson-Nyquist noise at the electron temperature, with no evidence of excess noise within our experimental resolution. The extracted $T_{el}$ is plotted as a function of $R_{total}$ in Fig. 3d. To interpret this dependence, we employ a minimal heat-balance model in which Joule heating is balanced by cooling through Wiedemann–Franz diffusion into the gold normal-metal electrodes and electron–phonon coupling [24], assuming that the phonon temperature equals the bath temperature — a valid approximation given the efficient thermal coupling of the gold electrodes to the substrate. The model prediction is shown as a solid curve, corresponding to the disordered limit of electron–phonon coupling in gold [23]. Here the base electron temperature at the bath, $T_{el}$ = 110 mK, is determined independently by quantum-dot thermometry (See Supplementary Information Fig. S3), providing a calibration point for the model without additional free parameters.

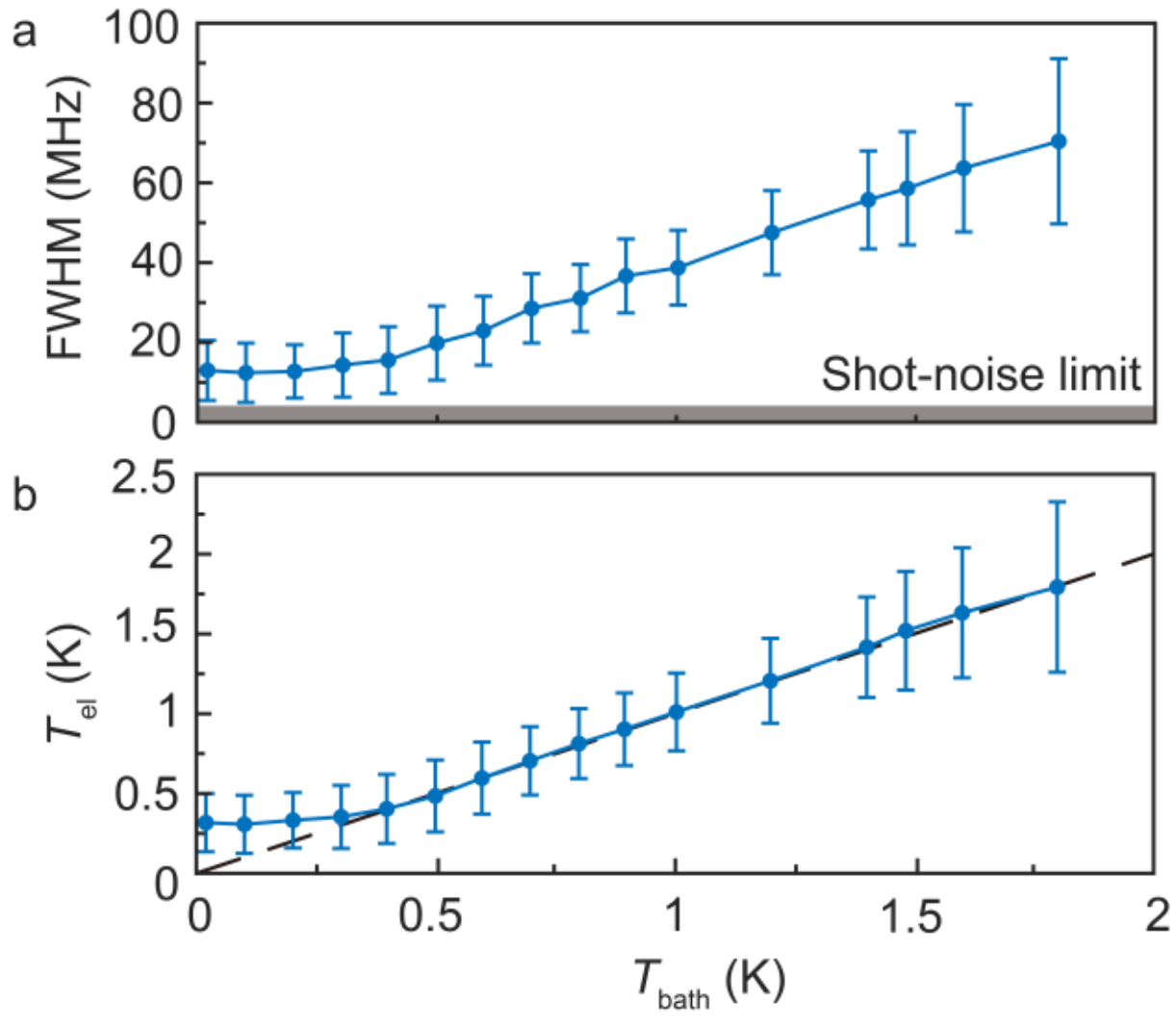

**Fig. 4 | Temperature dependence of linewidth and electron temperature.** (a) FWHM of the Josephson radiation as a function of bath temperature $T_{bath}$, measured at $V_{tg,JJ}$ = 2 V and $V_{tg,sh}$ = 3 V. Each data point represents the average FWHM of radiation peaks measured between 4.8 and 5.2 GHz, with error bars indicating the standard deviation. The shaded region indicates the shot-noise limit estimated from $I_b$. (b) $T_{el}$ extracted from the linewidth, plotted as a function of $T_{bath}$. The dashed line indicates $T_{el} = T_{bath}$.

## Electron thermometry under bias

To further validate the thermal noise interpretation of the linewidth, we examine its dependence on bath temperature $T_{bath}$. Figure 4a shows the FWHM as a function of $T_{bath}$, measured at $V_{tg,JJ}$ = 2 V and $V_{tg,sh}$ = 3 V. The linewidth increases monotonically with $T_{bath}$, consistent with the expected enhancement of Johnson-Nyquist noise at elevated temperatures [34,35]. Across the full temperature range, the measured linewidth remains well above the shot-noise limit estimated from $I_b$ [31,36,37], confirming that the emission linewidth is dominated by thermal noise rather than shot noise under our experimental conditions. Figure 4b shows $T_{el}$ extracted from the linewidth as a function of $T_{bath}$. At elevated temperatures, $T_{el}$ follows $T_{bath}$, indicating that the electronic system is well thermalized with the bath. At the lowest temperatures, however, $T_{el}$ saturates above $T_{bath}$, reflecting Joule heating by the bias current [24] that persists even at the lowest accessible bath temperatures. The saturation value is consistent with the $T_{el}$ extracted from the gate-dependent measurements in Fig. 3. The agreement of $T_{el}$ with the expected thermal behavior across the full range of $T_{bath}$ supports the interpretation that the linewidth provides a quantitative measure of the electron temperature and demonstrates that the Josephson radiation linewidth can serve as a direct probe of electron heating and cooling in this device [38].

## Conclusions

In summary, we have realized a graphene-based Josephson radiator in which the emission frequency and output power can be tuned largely independently using electrostatic gates. By combining a gate-tunable graphene Josephson junction with a separately controlled graphene shunt resistor, our device decouples the intrinsic coupling between frequency, power and dissipation that constrains conventional Josephson radiators, enabling independent control of both quantities on a single chip. Within this dual-gate operating space, we show that the linewidth of the emitted radiation is quantitatively accounted for by Johnson-Nyquist noise at the electron temperature, with no evidence of excess noise within our experimental resolution. Using the linewidth as a direct probe of the electron temperature, we further demonstrate that its dependence on the device resistance is captured by a minimal heat-balance model based on Wiedemann–Franz diffusion and electron–phonon coupling, without requiring additional cooling channels.

These results establish a direct and quantitative link between the spectral properties of Josephson radiation and electron thermal processes in a mesoscopic conductor. By connecting the emission linewidth to the electron temperature through Johnson-Nyquist noise, our approach provides a route to probe electron heating and cooling under bias, with relevance for nanoscale thermal devices and on-chip spectroscopic probes. From a device perspective, our work demonstrates an approach to controlling the frequency and power of Josephson radiation independently, achieved entirely with local electrostatic gates rather than an applied magnetic field [13]. Translating this platform into a practical in situ microwave source will require improvements along two directions. First, reducing the electron temperature through enhanced cooling — for instance, by increasing the volume of the normal-metal electrodes to enhance electron–phonon coupling — would directly narrow the emission linewidth. Second, engineering the total device resistance toward 50 Ω would improve impedance matching to standard microwave circuitry, maximizing the fraction of generated radiation delivered to the external load. Together, these refinements provide a clear pathway toward a gate-reconfigurable, low-noise on-chip Josephson source compatible with cryogenic quantum microwave circuits.

## Methods

**Sample Fabrication.** Monolayer graphene (MLG) and hexagonal boron nitride (hBN) flakes were mechanically exfoliated onto $Si/SiO_2$ substrates. Van der Waals heterostructures were assembled by dry transfer using polymer (Elvacite 2552C) stamps, with top and bottom hBN encapsulation layers each approximately 50 nm thick as determined by atomic force microscopy (AFM). The assembled stacks were then annealed in vacuum at 500 °C for 2 hours to remove residues and surface impurities, and clean regions free of bubbles were identified by AFM topography imaging prior to further processing. Electrodes and etch masks were patterned by standard electron-beam lithography. The hBN and graphene layers were subsequently etched by reactive ion etching (RIE) under $CF_4$ and $O_2$ plasma, respectively. For the shunt resistor, the graphene channel was patterned into an interdigitated electrode geometry to maximize the effective channel width. Normal-metal contacts to the shunt resistor were then formed by deposition of Ti/Au (5 nm / 70 nm) with a total contact area of 860 μm². Superconducting molybdenum–rhenium (MoRe, 100 nm) electrodes were subsequently deposited to contact both the graphene Josephson junction and the shunt resistor regions, connecting the two elements in parallel. Finally, a 20 nm gate dielectric layer of aluminum oxide ($AlO_x$) was deposited by atomic layer deposition (ALD) over the entire device, followed by patterning of MoRe top-gate electrodes for independent electrostatic control of the junction and shunt regions.

**DC Transport Measurement.** All measurements were performed in a dilution refrigerator at a base temperature of 20 mK. Gate voltages were applied using a voltage source (Yokogawa 7651). The graphene Josephson junction was current-biased by applying a voltage through a series resistance using the same voltage source. The junction voltage was amplified with a low-noise voltage amplifier (SP1004, Basel Precision Instruments) and recorded using a digital multimeter (Keithley 2000).

**Spectrum Analysis.** The power spectral density (PSD) of the emitted Josephson radiation was measured using a spectrum analyzer (N9010A, Keysight) with a resolution bandwidth (RBW) of 8 MHz, a video bandwidth (VBW) of 80 kHz, and 50 trace were averaged for each spectrum. The cryogenic microwave chain, schematically shown in Fig. 1a, consisted of a DC block (BLK-18-S+), bandpass filter (BPC21499, 4–8 GHz), lowpass filter (FLP-1250), isolator (LNF-ISISC4_12), and infrared filter (QMC-CRYOIRF-003MF), all thermally anchored at the mixing chamber stage (20 mK). The signal was transmitted from the device to the 4 K stage via a NbTi superconducting coaxial line, and from the 4 K stage to room temperature via a CuNi coaxial line. Cryogenic amplification was provided by a low-noise amplifier (LNF-LNC0.3_14B) at 4 K, followed by room-temperature amplifiers (ZX60-123LN-S+ and XR-A3Y2-0404C-01) connected in series. The usable measurement bandwidth was set by the bandpass filter to 4–8 GHz. Frequency-dependent background subtraction was performed by computing the ensemble-averaged PSD in the absence of Josephson radiation and subtracting it from the measured spectra in units of watts. To obtain the absolute radiation power, the measured power was further corrected for the net transmission of the RF measurement chain, characterized by de-embedding as described in Supplementary information Fig. S1.

**Acknowledgements**

This research was supported by the Samsung Science and Technology Foundation (Project No. SSTF-BA2401-03), the Agency for Defense Development through a grant funded by the Defense Acquisition Program Administration (DAPA) (UI257011TE), National Research Foundation of Korea (NRF) (nos., RS-2022-NR068223, RS-2024-00393599, RS-2024-00442710), and the ITRC (Information Technology Research Center) support program (IITP-2026-RS-2022-00164799). D.T.P. was partially supported by the NRF (no. RS-2024-00463743). Nanofabrication was partially supported by POSTECH Center for Semiconductor Technology Convergence supported by Korea Basic Science Institute, NFEC-2025-03-304835. K.W. and T.T. acknowledge support from the JSPS KAKENHI (Grant Numbers 21H05233 and 23H02052) and World Premier International Research Center Initiative (WPI), MEXT, Japan.

**Author contributions**

G.-H.L., J.K., and S.J. conceived the study. J.K. fabricated the devices with S.J.'s assistance. J.K. and S.L. built the low-noise measurement setup. J.K. performed the measurements. G.-H.L., J.K., S.J. and D.T.P. analyzed the data. K.W. and T.T. grew hBN crystals. G.-H.L. and S.J. supervised the work. G.-H.L., J.K. and D.T.P. wrote the manuscript. All authors reviewed and approved the manuscript.

**Competing interests**

The authors have no competing interests.

**Additional information**

Supplementary information. The online version contains supplementary material

available at https://doi.org./

Correspondence and requests for materials should be addressed to Gil-Ho Lee.

**References**

1 Krantz, P. *et al.* A quantum engineer's guide to superconducting qubits. *Appl. Phys. Rev.* **6**, 021318 (2019).

2 Aumentado, J. Superconducting Parametric Amplifiers: The State of the Art in Josephson Parametric Amplifiers. *IEEE Microw. Mag.* **21**, 45-59 (2020).

3 Hamilton, C. A. Josephson voltage standards. *Rev. Sci. Instrum.* **71**, 3611-3623 (2000).

4 Josephson, B. D. Possible new effects in superconductive tunnelling. *Phys. Lett.* **1**, 251-253 (1962).

5 Yanson, I., Svistunov, V. & Dmitrenko, I. Experimental observation of the tunnel effect for Cooper pairs with the emission of photons. *Sov. Phys. JETP* **21**, 650 (1965).

6 Langenberg, D. N., Scalapino, D. J. & Taylor, B. N. Josephson-type superconducting tunnel junctions as generators of microwave and submillimeter wave radiation. *Proc. IEEE* **54**, 560-575 (1966).

7 Welp, U., Kadowaki, K. & Kleiner, R. Superconducting emitters of THz radiation. *Nat. Photonics* **7**, 702-710 (2013).

8 Cassidy, M. C. *et al.* Demonstration of an ac Josephson junction laser. *Science* **355**, 939-942 (2017).

9 Varmazis, C., Lukens, J. E. & Finnegan, T. F. Broadband generation of tunable Josephson radiation at microwave frequencies. *Appl. Phys. Lett.* **30**, 660-661 (1977).

10 Koshelets, V. P. *et al.* Phase locked 270–440 GHz local oscillator based on flux flow in long Josephson tunnel junctions. *Rev. Sci. Instrum.* **71**, 289-293 (2000).

11 Mygind, J. *et al.* Phase locking of 270-440 GHz Josephson flux flow oscillators. *Supercond. Sci. Technol.* **12**, 720-722 (1999).

12 Lin, S.-Z. & Koshelev, A. E. Linewidth of the electromagnetic radiation from Josephson junctions near cavity resonances. *Phys. Rev. B* **87**, 214511 (2013).

13 Vervoort, S. *et al.* DC-operated Josephson junction arrays as a cryogenic on-chip microwave measurement platform. *Commun. Phys.* **8**, 292 (2025).

14 Deacon, R. S. *et al.* Josephson Radiation from Gapless Andreev Bound States in HgTe-Based Topological Junctions. *Phys. Rev. X* **7**, 021011 (2017).

15 Laroche, D. *et al.* Observation of the 4π-periodic Josephson effect in indium arsenide nanowires. *Nat. Commun.* **10**, 245 (2019).

16 Larson, T. F. Q. *et al.* Zero Crossing Steps and Anomalous Shapiro Maps in Graphene Josephson Junctions. *Nano Lett.* **20**, 6998-7003 (2020).

17 Takeshige, Y. *et al.* Experimental study of ac Josephson effect in gate-tunable $(Bi_{1-x}Sb_x)_2Te_3$ thin-film Josephson junctions. *Phys. Rev. B* **101**, 115410 (2020).

18 Haller, R. *et al.* ac Josephson effect in a gate-tunable $Cd_3As_2$ nanowire superconducting weak link. *Phys. Rev. B* **108**, 094514 (2023).

19 Heersche, H. B., Jarillo-Herrero, P., Oostinga, J. B., Vandersypen, L. M. K. & Morpurgo, A. F. Bipolar supercurrent in graphene. *Nature* **446**, 56-59 (2007).

20 Calado, V. E. *et al.* Ballistic Josephson junctions in edge-contacted graphene. *Nat. Nanotechnol.* **10**, 761-764 (2015).

21 Ben Shalom, M. *et al.* Quantum oscillations of the critical current and high-field superconducting proximity in ballistic graphene. *Nat. Phys.* **12**, 318-322 (2016).

22 Nanda, G. *et al.* Current-phase relation of ballistic graphene Josephson junctions. *Nano Lett.* **17**, 3396-3401 (2017).

23 Karvonen, J. T., Taskinen, L. J. & Maasilta, I. J. Electron–phonon interaction in thin copper and gold films. *Phys. Status Solidi C* **1**, 2799-2802 (2004).

24 Giazotto, F., Heikkilä, T. T., Luukanen, A., Savin, A. M. & Pekola, J. P. Opportunities for mesoscopics in thermometry and refrigeration: Physics and applications. *Rev. Mod. Phys.* **78**, 217-274 (2006).

25 Stewart, W. C. Current-voltage characteristics of Josephson junctions. *Appl. Phys. Lett.* **12**, 277-280 (1968).

26 McCumber, D. E. Effect of ac Impedance on dc Voltage-Current Characteristics of Superconductor Weak-Link Junctions. *J. Appl. Phys.* **39**, 3113-3118 (1968).

27 Golubov, A. A., Kupriyanov, M. Y. & Il'ichev, E. The current-phase relation in Josephson junctions. *Rev. Mod. Phys.* **76**, 411 (2004).

28 English, C. D. *et al.* Observation of nonsinusoidal current-phase relation in graphene Josephson junctions. *Phys. Rev. B* **94**, 115435 (2016).

29 Lee, G.-H., Kim, S., Jhi, S.-H. & Lee, H.-J. Ultimately short ballistic vertical graphene Josephson junctions. *Nat. Commun.* **6**, 6181 (2015).

30 Borzenets, I. V. *et al.* Ballistic Graphene Josephson Junctions from the Short to the Long Junction Regimes. *Phys. Rev. Lett.* **117**, 237002 (2016).

31 Likharev, K. K. *Dynamics of Josephson junctions and circuits*. (Gordon and Breach, 1986).

32 Stephen, M. J. Noise in the ac Josephson effect. *Phys. Rev.* **182**, 531 (1969).

33 Likharev, K. K. & Semenov, V. K. Fluctuation spectrum in superconducting point junctions. *JETP Lett.* **15**, 442 (1972).

34 Johnson, J. B. Thermal agitation of electricity in conductors. *Phys. Rev.* **32**, 97-109 (1928).

35 Nyquist, H. Thermal agitation of electric charge in conductors. *Phys. Rev.* **32**, 110 (1928).

36 Dahm, A. *et al.* Linewidth of the radiation emitted by a Josephson junction. *Phys. Rev. Lett.* **22**, 1416 (1969).

37 Schottky, W. Über spontane Stromschwankungen in verschiedenen Elektrizitätsleitern. *Ann. Phys.* **362**, 541-567 (1918).

38 Spietz, L., Lehnert, K. W., Siddiqi, I. & Schoelkopf, R. J. Primary Electronic Thermometry Using the Shot Noise of a Tunnel Junction. *Science* **300**, 1929-1932 (2003).